\documentclass[12pt]{iopart}

\usepackage{graphicx}

\begin{document}

\title{Probing the band structure of quadri-layer graphene with magneto-phonon resonance}

\author{C. Faugeras$^1$, P Kossacki$^1,2$, A.A.L. Nicolet$^1$, M. Orlita$^1$, M Potemski$^1$, A. Mahmood$^3$, D.M. Basko$^4$}
\address{$^1$ LNCMI-CNRS (UJF, UPS, INSA), BP 166, 38042 Grenoble Cedex 9, France}
\address{$^2$ Institute of Experimental Physics, University of Warsaw,Hoza 69, 00-681 Warsaw, Poland}
\address{$^3$ CNRS-Institut N$\acute{e}$el, BP 166, 38042 Grenoble Cedex 9, France}
\address{$^4$ Universit\'e Grenoble 1/CNRS, LPMMC UMR 5493, 25 rue des
Martyrs, 38042 Grenoble, France}

\ead{clement.faugeras@lncmi.cnrs.fr}

\date{\today}

\begin{abstract}
We show how the magneto-phonon resonance, particularly pronounced
in sp$^2$ carbon allotropes, can be used as a tool to probe the
band structure of multilayer graphene specimens. Even when
electronic excitations cannot be directly observed, their coupling
to the E$_{2g}$ phonon leads to pronounced oscillations of the
phonon feature observed through Raman scattering experiments with
multiple periods and amplitudes detemined by the electronic
excitation spectrum. Such experiment and analysis have been
performed up to 28T on an exfoliated 4-layer graphene specimen
deposited on SiO$_2$, and the observed oscillations correspond to
the specific AB stacked 4-layer graphene electronic excitation
spectrum.

\end{abstract}
\submitto{\NJP}

\maketitle

\section{Introduction}

Since the theoretical prediction of magneto-phonon resonance in
graphene and bilayer
graphene~\cite{Ando2007a,Goerbig2007,Ando2007b}, the resonant
interaction between inter-Landau level electronic excitations and
$\Gamma$-point optical phonon in graphene, this effect has been
widely explored experimentally in neutral graphene-like systems
such as multi layer epitaxial graphene on C-face
SiC~\cite{Faugeras2009}, graphene like inclusions on the surface
of bulk graphite~\cite{Yan2010,Faugeras2011}, and in bulk
graphite~\cite{Kossacki2011,Kim2012}. Magneto-phonon resonance in
these systems are pronounced because (i) they are gapless and low
energy direct ($\Delta k =0$) electronic excitations exist, and
(ii) because of the Kohn anomaly~\cite{Kohn1959,Piscanec2004} at
the $\Gamma$ point of the phonon Brillouin zone which makes phonon
energies \textit{ultra sensitive} to modifications of the
electronic excitation spectrum. Such modifications can be achieved
by tuning the Fermi energy or by applying a magnetic field
perpendicular to the plane of the quasi-2D crystal.

\begin{figure}
\includegraphics[width=1\textwidth]{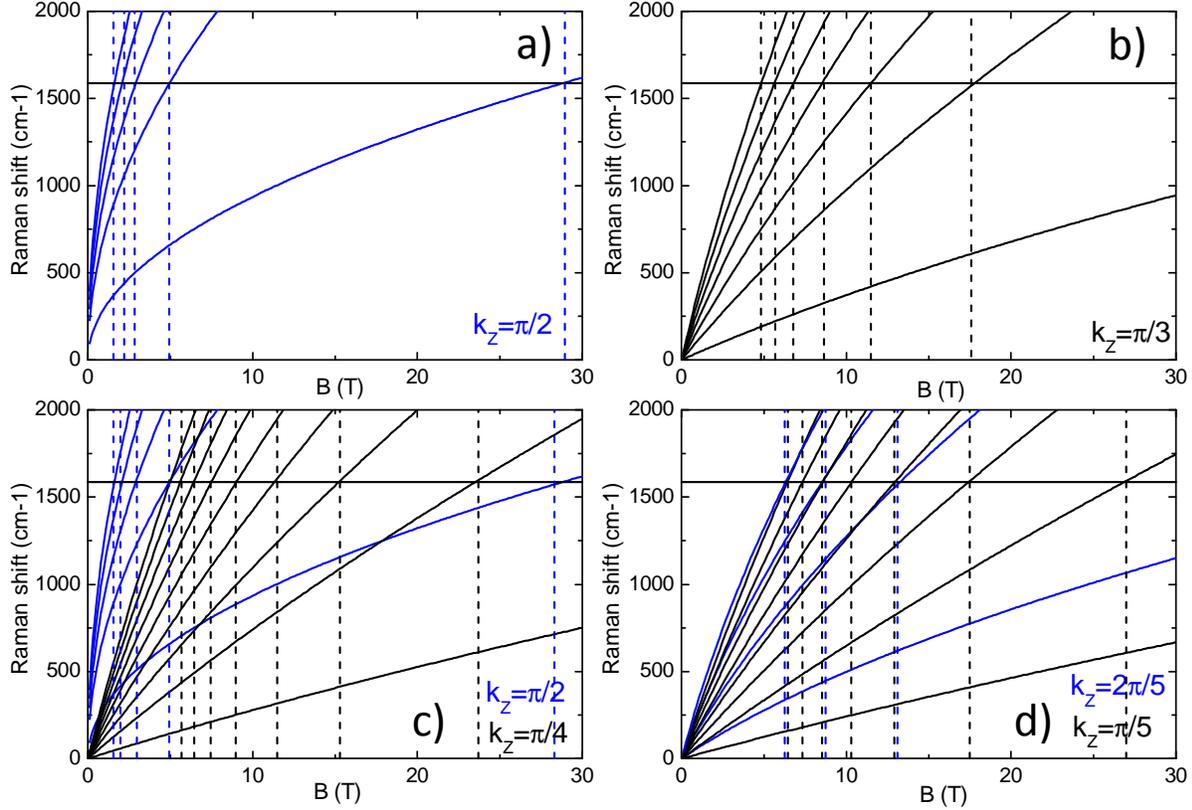}
\caption{\label{Fig1} Evolution of inter Landau level excitations
with $\Delta |n|=\pm1$ calculated in the frame of the model
developed in Ref.~\cite{Koshino2008}, in the case of (a) a
monolayer graphene, (b) a bilayer graphene, (c) a trilayer
graphene and (d) four-layer graphene. The values of $k_z$ used in
these calculations are indicated in the figures. The phonon energy
is indicated by the solid horizontal line at 1585~cm$^{-1}$.
Dashed vertical lines indicate resonant magnetic fields
corresponding to the different excitation spectra. Colors are only
used to differentiate the different excitation spectra.}
\end{figure}

Performed in systems with a well-known electronic excitation
spectrum, magneto-phonon resonance experiments provide precise
information concerning the electron-phonon interaction through the
amplitude of the observed oscillations of the energy and of the
line width of the phonon feature observed in Raman scattering
experiments, as a function of the applied magnetic field. On the
other hand, since electronic excitations which couple to the
E$_{2g}$ optical phonons are well identified by symmetry arguments
(optical-like excitations characterized by a change of angular
momentum $\Delta |n|=\pm 1$~\cite{Ando2007a}, where $n$ is the
Landau level index), this effect can be used to determine the
electronic excitation spectrum. The evolution of optical-like
excitations, calculated in the frame of Ref.~\cite{Koshino2008},
as a function of the magnetic field is presented in
figure~\ref{Fig1} for a) graphene, b) bilayer graphene, c)
tri-layer graphene and d) for four-layer graphene. Each of these
distinct electronic excitation spectra are tuned in resonance with
the phonon energy at different characteristic values of the
magnetic field. In analogy to magneto-transport experiments which
probe the density of states at the Fermi energy, magneto-phonon
resonance can be used as a probe of $\Delta |n|=\pm 1$ electronic
excitation at the E$_{2g}$ phonon energy, and Landau level
spectroscopy can be achieved by modelling the observed shifts of
the phonon feature.

In this paper, we report on low-temperature magneto-Raman
scattering experiments revealing the magneto-phonon resonance of
an exfoliated multi-layer graphene specimen deposited on a
SiO$_2$/Si substrate. The investigated sample is characterized by
a low ``intrinsic'' doping and oscillations of both the phonon
energy and line width can be observed for magnetic fields
$B>7\:\mbox{T}$. These oscillations have a noticeably more
complicated pattern than for monolayer graphene and for bulk
graphite. From the analysis of this pattern we deduce the number
of layers in the sample, which turns out to be equal to four.
Observation of a magneto-phonon resonance can hence be used to
trace electronic excitations at the phonon energy and determine
the band structure of the probed electronic system.

\section{Experimental details}

\begin{figure}
\includegraphics[width=1\textwidth]{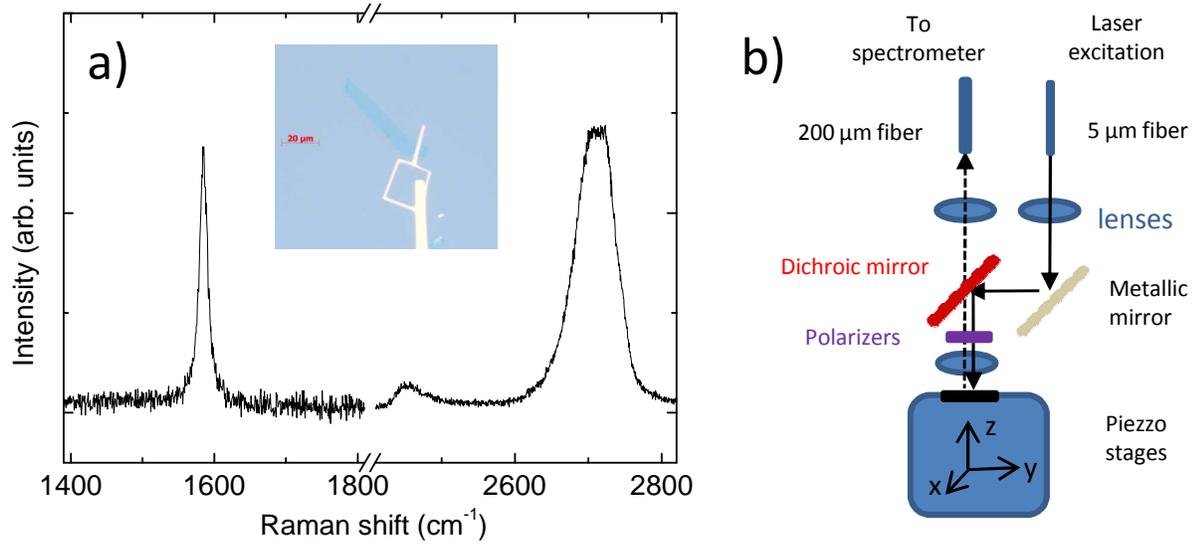}
\caption{\label{Fig2} a) Raman scattering spectrum of the
four-layer graphene specimen measured at room temperature with a
LABRAM confocal micro-Raman scattering set-up, with the
$\lambda=514.5$ nm line of an argon laser. Inset: Microscope image
of the 4-layer graphene flake. b) Schematics of the micro-Raman
scattering set-up operating at liquid helium temperature and in
high magnetic fields.}
\end{figure}

The multilayer graphene specimen is presented in the inset of
figure~\ref{Fig2}a. It has been exfoliated from bulk graphite and
deposited on a Si/SiO$_2$ substrate with $300$ nm of SiO$_2$. A
single metallic electrode was then deposited on the flake (not
used in this experiment). To locate the sample, we spatially map
the Raman scattering signal of the silicon substrate, which shows
a strong attenuation when the laser spot is focused on the
metallic layer. A typical room temperature Raman scattering
spectrum of the multilayer graphene flake measured with LABRAM
confocal micro-Raman scattering set-up, together with an optical
image are presented in figure~\ref{Fig2}a. The 2D band has a
multi-components broad feature. We find the standard analysis of
2D band features in multi-layers specimens, that is, the
deconvolution of the complex 2D band line shape using $2N$
Lorentzian functions (where $N>1$ is the number of layers),
relevant for mono-layers, for bi-layers and for
tri-layers~\cite{Ferrari2006,Graf2007,Lui2011}, not precise enough
for $N>3$. As a consequence, 4-layer graphene flakes can hardly be
distinguished from the 5-layer and additional information is
needed to fully characterized multilayer graphene systems.

The low-temperature magneto-Raman scattering response of our
multilayer graphene has been measured with a home-made micro-Raman
scattering setup presented in figure~\ref{Fig2}b. Laser excitation
at $\lambda=488\:\mbox{nm}$ produced by an Ar ion laser is
injected into a $5\:\mu\mbox{m}$ core mono-mode optical fiber
which is fixed on a miniaturized optical bench and placed at the
center of the magnet. This experimental setup can host various
optical filters which are used to filter the excitation laser
light at the output of the mono mode fiber (laser line filter), to
selectively reflect the excitation laser and transmit the
scattered signal (dichroic mirror), to reject the scattered laser
before the collection fiber (notch filter), and to impose a
circular polarization (linear polarizers and quarter wave plate).
The excitation spot on the surface of the sample has a diameter of
$\sim{1}\:\mu\mbox{m}$ and the optical power is of 5~mW. The
sample is mounted on piezzo stages which allow to move the sample
under the laser spot with a sub-micrometer resolution. Polarized
Raman scattering spectra were measured in nearly back-scattering
Faraday geometry. The collected light was dispersed with a single
grating spectrometer (spectral resolution
$\Delta\lambda=0.07\:\mbox{nm}$) equipped with nitrogen cooled CCD
detector and band pass filters were used to reject the stray
light.

\section{Magneto-Raman scattering of exfoliated multilayer graphene specimen}

In order to obtain further information concerning our multilayer
graphene sample, we have performed magneto-Raman scattering
experiments in the G-band range of energy. The G-band feature of
the Raman scattering spectrum is observable in the crossed
circular polarization configuration~\cite{Faugeras2011} which is
used in the present experiment. Even though electronic excitations
cannot be, in the present case, directly imaged on the Raman
scattering spectra, their effect when tuned in resonance to the
E$_{2g}$ optical phonon by applying a magnetic field can be
clearly observed.

\begin{figure}
\includegraphics[width=1\textwidth]{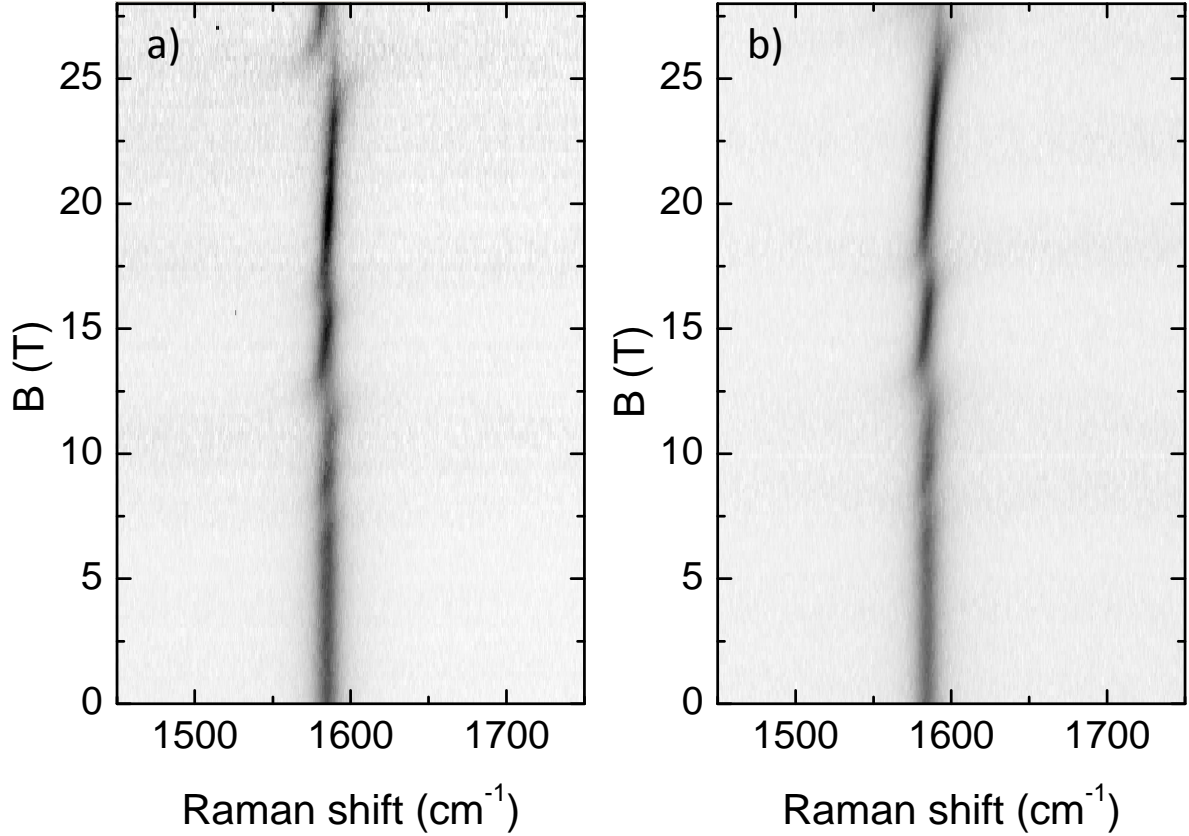}
\caption{\label{Fig3} False color map of the scattered intensity
in the energy range of the $E_{2g}$ phonon as a function of the
magnetic field in the $\sigma^-/\sigma^+$ (a) and
$\sigma^+/\sigma^-$ (b) polarization configurations.}
\end{figure}

\begin{figure}
\includegraphics[width=1\textwidth]{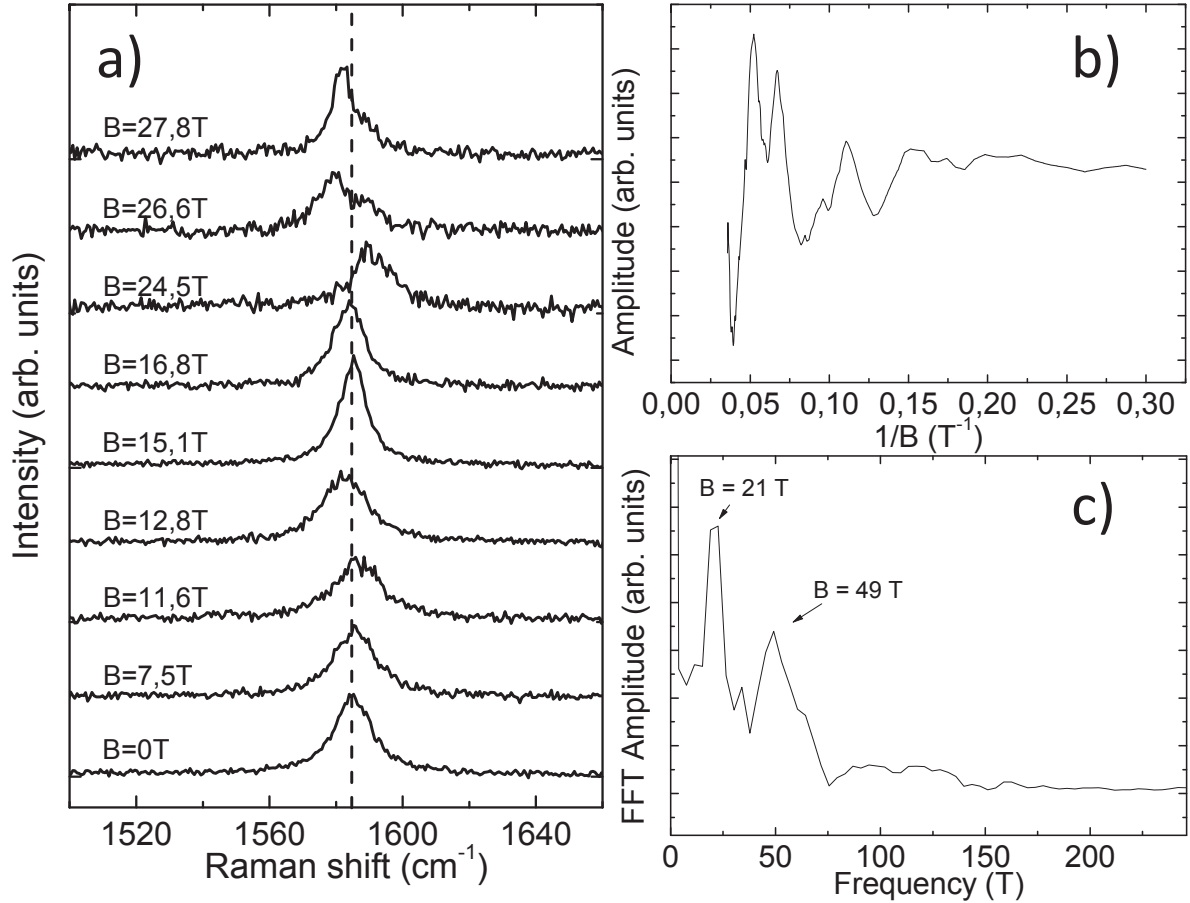}
\caption{\label{Fig4} a) Raman scattering spectra measured in the
$\sigma^-/\sigma^+$ polarization configuration at different values
of the magnetic field. The dashed vertical line indicates the B=0
energy of the $E_{2g}$ phonon. b) Intensity of the scattered light
at the phonon energy at B=0 as a function 1/B. c) Amplitude of the
Fourier transform of b) showing two characteristic frequencies.}
\end{figure}

The magneto-phonon resonance of our multilayer graphene sample is
presented in figure~\ref{Fig3} in the form of false color map of
the scattered intensity as a function of the magnetic field, in
the two crossed circular polarization configurations
$\sigma^{\pm}/\sigma^{\mp}$. Individual spectra at characteristic
values of magnetic field in $\sigma^{-}/\sigma^{+}$ configuration
are presented in figure~\ref{Fig4}a. In both configurations, we
observe, for $B>7\:\mbox{T}$, oscillations of the energy of the
$E_{2g}$ phonon and of its line width. This low value of the
magnetic field for the onset of magneto-oscillations is an
indication that the carrier density in this sample is lower than
half of the $E_{2g}$ phonon energy. The resonant values of the
magnetic field are different in the two polarization
configurations,
$B_{res}=7,\textbf{7.85},10,\textbf{11.9},16.75,26.6\:\mbox{T}$ in
the $\sigma^{+}/\sigma^{-}$ configuration and
$B_{res}=6.7,\textbf{7.3},9.7,\textbf{11.4},15.7,25.6\:\mbox{T}$
in the $\sigma^{-}/\sigma^{+}$ configuration. Moreover, the
electron-phonon coupling appears to be more efficient at
particular values of the magnetic field (indicated above by the
bold typeface) where the amplitude of the observed oscillation is
enhanced with respect to other resonant magnetic fields. This
nontrivial behavior is different from the one observed for
epitaxial graphene, for graphene-like domains on the surface of
bulk graphite or for K-point carriers in bulk graphite, where the
amplitude of the oscillations was found to increase with the
magnetic field as the effective electron-phonon coupling strength
depends on the degeneracy of the Landau levels which increases
linearly with~$B$.

To determine the number of Landau level fan charts contributing to
the observed magneto-phonon resonance, we extract from the spectra
presented in figure~\ref{Fig3}a and in figure~\ref{Fig4}a the
evolution of the amplitude of the scattered light at the phonon
energy at $B=0$ as a function of the inverse magnetic field. This
quantity shows oscillations which are representative of the phonon
oscillations: each time the phonon feature splits or is shifted
due to the resonant interaction with an electronic excitation, the
amplitude of the scattered light at the phonon energy at $B=0$
decreases. This evolution is presented in figure~\ref{Fig4}b. It
is then possible to numerically calculate the Fourier transform of
this signal, presented in figure~\ref{Fig4}c, which shows two
distinct frequencies at about 21 and 49~T, corresponding to two
periods $\Delta(1/B)$ of about $0.05\:\mbox{T}^{-1}$ and
$0.02\:\mbox{T}^{-1}$.

Their appearance can be understood qualitatively using the results
of Ref.~\cite{Koshino2008} for the electronic spectrum of a
$N$-layer graphene. Namely, the spectrum is similar to that of a
bulk graphite, but instead of the continuous quantum number~$k_z$,
$-\pi/2<k_z\leq\pi/2$ (the wave vector, perpendicular to the
layers, in the units of the inverse interlayer spacing), only
$N/2$ (for $N$ even) or $(N+1)/2$ (for $N$ odd) discrete values,
$k_z=\pi/(N+1),\ldots,(N/2)\pi/(N+1)$ or
$k_z=\pi/(N+1),2\pi/(N+1),\ldots,\pi/2$, respectively, are
allowed. Now, instead of a continuum of inter-Landau-level
excitations for all~$k_z$ in graphite, which couple to the
phonon~\cite{Kossacki2011}, in the 4-layer graphene one has two
series of inter-Landau-level excitations, corresponding to the two
allowed values of~$k_z$. Thus, the presence of two periods
indicated that our sample has either three or four layers.

In the two-band approximation for the effective bilayer model of
Ref.~\cite{Koshino2008}, the values of the magnetic field $B_n$ at
which the magnetophonon resonance occurs, are determined by
\begin{equation}
\left(\sqrt{n(n-1)}+\sqrt{n(n+1)}\right)\frac{v_{F}^{2}eB_n}{\gamma_1\cos{k}_z}
=\omega_\mathrm{ph},
\end{equation}
where $v_{F}$~is the Fermi velocity in the monolayer, and
$\gamma_1$ is the nearest-neighbor inter-layer matrix element.
Approximating $\sqrt{n(n\pm{1})}\approx{n}\pm{1}/2$, we obtain
\begin{equation}
\frac{1}{B_n}=\frac{2v_{F}^2e}{\omega_\mathrm{ph}\gamma_1\cos{k}_z}\,n.
\end{equation}
The coefficient in front of~$n$ thus gives the period $\Delta(1/B)$
for each value of~$k_z$. For the monolayer-type spectrum at
$k_z=\pi/2$, the same consideration (again, approximating
$\sqrt{n(n\pm{1})}\approx{n}\pm{1}/2$) gives
\begin{equation}
\frac{1}{B_n}=\frac{8v_{F}^2(e/c)}{\omega_\mathrm{ph}^2}\,(n+1/2).
\end{equation}
For $\omega_\mathrm{ph}=196\:\mbox{meV}$ and
$v_{F}=1.01\times{10}^6\:\mbox{m}/s$ solving this equation brings
an inverse period of 7.1~T, which is too small compared to the
smallest observed 21~T. Thus, we exclude the possibility that our
sample could be a trilayer graphene. For four layers, we have
$k_z=\pi/5,2\pi/5$, which, for $\gamma_1=380\:\mbox{meV}$, gives
two frequencies of 16 and 41~T. This is an indication that the
sample has four layers.

\begin{figure}
\includegraphics[width=1\textwidth]{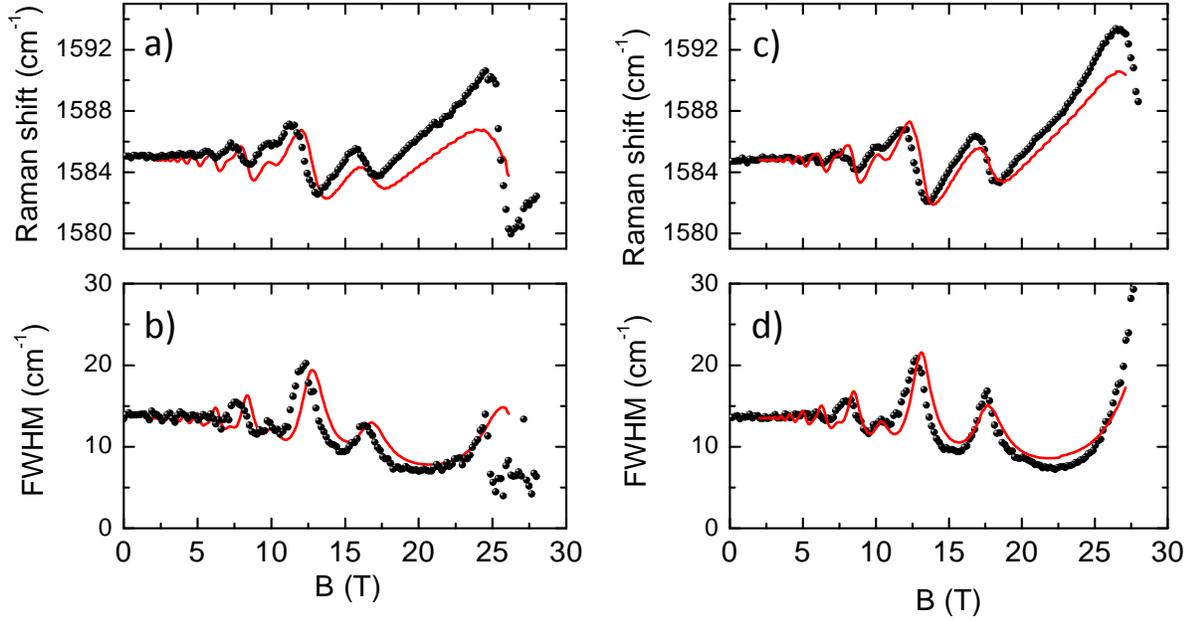}
\caption{\label{Fig5}Evolution of the maximum of scattered light
and of the FWHM of the G band as a function of the magnetic field
in the $\sigma^-/\sigma^+$ (a and b) and in the
$\sigma^+/\sigma^-$ polarization configurations. Red solid lines
are the result of the calculation detailed in the text.}
\end{figure}

To give more solid arguments, we perform a quantitative modelling
of the spectra. The Raman scattering spectra presented in
figure~\ref{Fig4} can be described by Lorentzian functions to
extract the energy corresponding to the maximum of scattered light
and the full width at half maximum (FWHM) of the phonon feature.
These results are presented by the black dots in figure~\ref{Fig5}
(a-d) for both polarization configurations. Again, the observed
behavior is not a simple series of oscillations; rather, two
series of oscillations can be distinguished. Qualitatively, this
is explained by the same arguments as above (two allowed values
of~$k_z$). The theoretical curves are shown in figure.~\ref{Fig5}
as solid curves on top of the experimental symbols.

\section{Theoretical aspects of the magneto-phonon resonance
in 4-layer graphene}

In contrast to bulk graphite, where due to the translational
invariance in the $z$~direction, the momentum conservation holds,
which implies that only phonons with the wave vector $q_z=0$ are
probed (assuming the photon momentum to be very small), and that
the electronic transitions involved are those from states with
some $k_z$ into states with $k_z'=k_z-q_z=k_z$, the 4-layer
graphene lacks the translational invariance in the $z$~direction,
which introduces some additional difficulties in the theoretical
description of the magneto-phonon resonance, as compared to the
bulk graphite.

(i)~There is \emph{a priori} no reason to focus on one particular
phonon mode. In total, there are eight degenerate optical phonon
modes with zero in-plane wave vector (two for each layer). The
circular basis for the in-plane polarizations is still a good one,
thus for each circular polarization there are four phonon modes.
We assume that their degeneracy is lifted by the electron-phonon
interaction, neglecting the elastic coupling between the layers.
Thus, for each of the two circular polarizations, the normal modes
should be found as eigenvectors of the electronic polarization
operator $\Pi(\omega)$, which is a $4\times{4}$ matrix.

(ii)~The electronic polarization operator not only has a
contribution from the $k_z$-diagonal electronic transitions, but
also from transitions between states with different~$k_z$, since
no momentum conservation holds. This gives not just two series of
transitions, but four, corresponding to all possible combinations
of $\{\pi/5,2\pi/5\}\to\{\pi/5,2\pi/5\}$. Still, it can be
naturally expected (and is confirmed by the actual calculation)
that the overlap between states with different~$k_z$ is much
smaller than that between states with the same~$k_z$; this is why
two series of oscillations instead of four are seen in
figure~\ref{Fig4}.

(iii)~Generally speaking, all four phonon modes should contribute
to the Raman spectrum. The intensities of the four components are
determined by the projections of the corresponding mode
eigenvectors on the Raman matrix element (photon-photon-phonon
vertex), which is also a four-component vector for each given
circular polarization configuration. In bulk graphite, due to the
momentum conservation, the Raman vertex would be simply
proportional to a $\delta$-function $\delta(q_z)$, but in the
4-layer sample it has a non-trivial structure.

The electronic band structure of multilayer graphene is modelled
similarly to Koshino and Ando~\cite{Koshino2008}. Namely, in the
nearest-layer tight-binding approximation, the band structure of
any Bernal-stacked multi-layer graphene can be deduced from the
one of the bulk graphite by selecting specific values of~$k_z$
corresponding to the possible standing waves perpendicular to the
basal plane of the multilayer graphene specimen. The monolayer
graphene then corresponds to a single value of $k_z=\pi/2$, the
bilayer to $k_z=\pi/3$, the trilayer corresponds to two values of
$k_z=\pi/4,\pi/2$, etc. At each value of $k_z$, the Hamiltonian
has the form of an effective graphene bilayer:
\begin{equation}\label{bilayerHam=}
{H}_{k_z}(\hat{\vec{p}})=\left[\begin{array}{cccc} \Gamma_2 &
v_F\hat{p}_- & -\alpha_4v_F\hat{p}_- & 0 \\
v_F\hat{p}_+ & \Gamma_5 & \Gamma_1 & -\alpha_4v_F\hat{p}_- \\
-\alpha_4v_F\hat{p}_+ & \Gamma_1 & \Gamma_5 & v_F\hat{p}_-\\
0 & -\alpha_4v_F\hat{p}_+ & v_F\hat{p}_+ & \Gamma_2
\end{array}\right],
\end{equation}
where
\begin{equation}
v_F=\frac{3}{2}\,\frac{\gamma_0{a}}{\hbar},\;\;\;
\Gamma_1=2\gamma_1\mathcal{C},\;\;\;
\Gamma_2=2\gamma_2\mathcal{C}^2,\;\;\;
\alpha_{4}=\frac{2\gamma_{4}}{\gamma_0}\,\mathcal{C},\;\;\;
\Gamma_5=2\gamma_5\mathcal{C}^2+\Delta,
\end{equation}
$\mathcal{C}\equiv\cos{k}_z$, $a=1.42\:\mbox{\AA}$ is the distance
between the neighboring carbon atoms in the same layer,
$\hat{p}_\pm=-i\hbar(\partial_x\pm{i}\partial_y)$ are the in-plane
momenta counted from the corners of the hexagonal first Brillouin
zone, and $\gamma_0=3.1\:\mbox{eV}$, $\gamma_1=0.38\:\mbox{eV}$,
$\gamma_2=-20\:\mbox{meV}$, $\gamma_4=44\:\mbox{meV}$,
$\gamma_5=94\:\mbox{meV}$, $\Delta=-8\:\mbox{meV}$ are the
parameters of the tight-binding Slonczewski-Weiss-McClure (SWM)
model~\cite{SWM1,SWM2}, whose values we take the same as in
Ref.~\cite{Kossacki2011}. As in Ref.~\cite{Kossacki2011}, we have
neglected the $\gamma_3$~term, responsible for the trigonal
warping, in order to simplify the diagonalization of the
electronic Hamiltonian with an external magnetic field. The
$\gamma_2,\gamma_5$ terms in the SWM model correspond to the
second-nearest-layer coupling, and their non zero values modify
the electronic wave functions on the layers with respect to the
simple recipe of Ref.~\cite{Koshino2008}. Here, we consider the
wave functions of Ref.~\cite{Koshino2008} but we still include the
$\gamma_2$ and $\gamma_5$ terms in Eq.~(\ref{bilayerHam=}), since
the experimental transition energies are better reproduced when
also considering these terms. Thus, even though
Eq.~(\ref{bilayerHam=}), strictly speaking, does not correspond to
the SWM model, we use it as an effective electronic Hamiltonian
for the 4-layer graphene. Given Eq.~(\ref{bilayerHam=}), the
calculation of the polarization operator is analogous to the one
presented in Ref.~\cite{Kossacki2011} with summation over
$k_z,k_z'$ instead of integration over~$k_z$. Following points (i)
and (ii) discussed at the beginning of this section, we also
introduce a layer index for the electron-phonon coupling. The
solid curves presented in figure.~\ref{Fig5} were obtained using
the values $\lambda=4\times{10}^{-3}$, $\delta=100\:\mbox{meV}$
for the dimensionless electron-phonon coupling constant and the
electronic broadening respectively, but the values
$\lambda=3\times{10}^{-3}$, $\delta=80\:\mbox{meV}$ or
$\lambda=5\times{10}^{-3}$, $\delta=120\:\mbox{meV}$ differ very
little. (We remind that the main purpose of the present modelling
is not to extract precise values of parameters but to demonstrate
that the sample is quadri-layer graphene.)

The third difficulty, mentioned at the beginning of this section,
can be bypassed since we are interested only in the positions and
widths of the Raman peaks, determined by the complex eigenvalues
of the electronic polarization operator $\Pi(\omega)$, i.~e., by
the complex roots of the equation
\begin{equation}
\det\left[\omega^2-\omega_0^2-2\omega_0\Pi(\omega)\right]=0,
\end{equation}
and not in their intensities. Moreover, it turns out that the four
eigenvalues behave quite differently upon varying the magnetic
field. Only one eigenvalue exhibits pronounced
magneto-oscillations, while other three remain close to the bare
phonon frequency. This behavior can be understood by noting that
near each resonance~$\alpha$, the polarization operator can be
approximated as $\Pi_{ij}(\omega)\sim
v_i^\alpha(v_j^\alpha)^*/(\omega-\omega_\alpha+i\Gamma_\alpha)$,
where $\omega_\alpha$ and $\Gamma_\alpha$ are the frequency and
the broadening of the corresponding transition, and $v_i^\alpha$
is the corresponding matrix element of the electron-phonon
coupling (the indices $i,j=1,2,3,4$ label the layers where the
phonon displacements occur). Then, near the resonance, one phonon
eigenvector is along $v^\alpha_i$, and other three are orthogonal
to it, so that they are effectively decoupled from the resonance.
The theoretical curves in figure~\ref{Fig5} correspond to the
eigenvalue exhibiting the non-trivial oscillating behavior.
Nevertheless, the contribution of other modes, not subject to
oscillations, can also be distinguished in the spectrum of
Fig.~\ref{Fig3}a) around the strongest resonance at
$B=25\:\mbox{T}$. Finally, we note that the possibility to avoid
the calculation of the Raman matrix element in the 4-layer
structure is quite fortunate, since its calculation requires a
significant effort even in the monolayer
graphene~\cite{Basko2009}.

\section{Conclusions}

We have shown that magneto-phonon resonance in multi-layer
graphene specimens can be used to determine the specific band
structure of these systems. We demonstrate this possibility in
this work with a 4-layer graphene flake deposited on SiO$_2$.
Magneto-phonon resonance in this sample result in a series of
oscillations of the phonon energy which can be directly related to
resonances between two distinct Landau Levels fan charts,
characteristic of the 4-layer graphene system.

\ack Part of this work has been supported by EC
MTKD-CT-2005-029671, EuroMagNET II under the EU contract number
228043 and PICS-4340 projects. P.K. is financially supported by
the EU under FP7, contract no. 221515 `MOCNA'.

\end{document}